\documentstyle[aps,epsf,epsfig,preprint]{revtex}

\def\prl#1#2#3{{ Phys. Rev. Lett.} {\bf #1}, #2 (#3)}

\def\pla#1#2#3{Phys. Lett. A {\bf #1}, #2 (#3)}
\def\pre#1#2#3{Phys. Rev. E {\bf #1}, #2 (#3)}
\def\prb#1#2#3{Phys. Rev. B {\bf #1}, #2 (#3)}
\def\prlon#1#2#3{Proc. Phys. Soc. London A {\bf #1}, #2 (#3)}
\def\prp#1#2#3{Phys. Rep. {\bf #1}, #2 (#3)}
\def\cmp#1#2#3{Comm. Math. Phys.  {\bf #1}, #2 (#3)}
\def\fund#1#2#3{{\it Fund. Math. }{\bf #1}, #2  (#3) }

\def\lanl{{\it LANL archives}}
\def\jsph#1#2#3{Journal of Statistical Physics {\bf #1}, #2 (#3)}

\def\pr#1#2#3{Phys. Rev. {\bf #1}, #2 (#3)}

\def\jpha#1#2#3{J. Phys. A, Math Jr. {\bf #1}, #2 (#3)}
\def\jpcs#1#2#3{J. Phys. C. Solid State Physics {\bf #1}, #2 (#3)}

\def\physd#1#2#3{Physica D {\bf #1}, #2 (#3)}
\def\pram#1#2#3{Pramana J. Phys. {\bf #1}, #2 (#3)}

\def\etl{$et ~al.$~}

\def\beqr{\begin{eqnarray}}
\def\eqnr{\end{eqnarray}}
\def\beq{\begin{equation}}
\def\bc{\begin{center}}
\def\ec{\end{center}}
\def\eqn{\end{equation}}
\begin{document}
\title{\huge{Critical Localization and Strange Nonchaotic Dynamics:
The Fibonacci Chain}}
\author{Surendra Singh Negi and Ramakrishna Ramaswamy} 
\address{School of Physical Sciences\\ Jawaharlal Nehru University,
New Delhi 110 067, INDIA} \date{\today} \maketitle
\begin{abstract}
The discrete Schr\"odinger equation with a quasiperiodic dichotomous
potential specified by the Fibonacci sequence is known to have a
singular continuous eigenvalue spectrum with all states being
critically localized. This equation can be transformed into a
quasiperiodic skew product dynamical system. In this iterative
mapping which is entirely equivalent to the Schr\"odinger problem,
critically localized states correspond to fractal attractors which
have all Lyapunov exponents equal to zero. This provides an alternate
means of studying the spectrum, as has been done earlier for the
Harper equation. We study the spectrum of the
Fibonacci system and describe the scaling of gap widths with
potential strength.
\end{abstract}
\section{INTRODUCTION}

The phenomenon of localization, first discussed in the context of
disordered systems by Anderson \cite{anderson}, has been of great
interest in condensed matter physics for several decades. The
Anderson model \cite{anderson,bds,hike,mott-twose} 
is the tight--binding Hamiltonian
\beq
H =\sum_{i} W_i c_i^{\dag} c_i + \sum_{i,j}I_{ij} c_i^{\dag} c_j, 
\eqn
where $c_{i}^{\dag}$ is the creation operator at the lattice
site indexed by $i$, and $W_i$ and
$I_{ij}$ are on--site and hopping matrix elements. In the original
model, $I_{ij}= I$ for $i,j$ nearest neighbours and zero otherwise,
and $W_{i}$ is chosen randomly in the interval [$-W/2, W/2$].  $I$
and $W$ can be combined into a dimensionless ratio which is a measure
of the disorder.  For $\frac{W}{I} > [\frac{W}{I}]_{c}$, a critical
value, all the states are exponentially localized. Below the critical
value, there is a mobility edge and states can be extended.  For
ordered systems, such as the analogous system with a periodic
potential, states are extended and the energy spectrum is a band
\cite{Azbel}.  

There are systems that lie between these extremes, for instance crystals
with additional periodic modulation which is incommensurate with the
period of the underlying lattice \cite{soko,thpb,aubry}. The lack of
strict translational order has an effect similar to disorder, and
thus it is of interest to know the nature of the states in such
systems. In this paper we study the spectrum of a 1--d chain, the
potential at each site being determined by a quasiperiodic sequence
deriving from the Fibonacci numbers \cite{fib11,fib12,fib13,fib14}. 

The paradigm for studies of quasiperiodic Schr\"odinger problems has
been the Harper equation \cite{Harper} 
\beq
\psi_{n+1} + \psi_{n-1} + V_n \psi_n = E \psi_n
\label{harpq}, 
\eqn
which is the discrete Schr\"odinger equation for a particle in the
quasiperiodic potential $ V_n = 2 \epsilon \cos 2 \pi(n\omega+
\phi_0)$, with $\omega$ being irrational. This is a tight--binding model
describing the motion of a electron in two dimensional lattice in the
presence of a magnetic field, but is also known to arise in numerous
other contexts. As a result, it has been extensively studied both in the
physics \cite{thouless,barelli,kohmoto,pandit,kohmoto2,luck,ag,gisel} as well as
 in the mathematics \cite{sinai,simon,bell,john,numbther}
literature (as the ``almost--Mathieu'' equation).  Since the results
that are known for this system have considerable bearing on the
present work, we first briefly review the salient features of
the Harper system.

Transfer matrices provide a powerful methodology for the calculation
of the band spectrum \cite{stockmann} in matrix form,
the discrete Schr\"odinger equation, Eq.(\ref{harpq}) can be written
as
\beq
\left ( \begin{array}{c}
\psi_{n+1}\\
\psi_{n}
\end{array} \right) = T_{n} \left ( \begin{array}{c}
\psi_{n}\\
\psi_{n+1} \end{array} \right) 
\eqn
where the transfer matrix $T_{n}$ is given by
\begin{equation}
T_{n}=\left(\begin{array}{clcl} E-V_n & -1 \\ 1 & 0
\end{array}\right)
\end{equation}
For 1--d chains, the state vector at the final site is obtained from the
starting value by matrix multiplication,
\beq
\left ( \begin{array}{clcr}
\psi_{N+1}\\
\psi_{N}
\end{array} \right)  = T \left (
\begin{array}{clcr}
\psi_{1}\\
\psi_{0}
\end{array} \right) \eqn
with $T = T_{N} \cdot T_{N-1} \cdot T_{N-2} \cdots T_{1}$. The
eigenvalues of $T_{n}$ can be calculated by the Cayley-Hamilton
theorem as 
\beq
\lambda =\frac{E-V_n \pm \sqrt{(E- V_n)^2-4}}{2}.
\eqn
From this equation it is evident that $\lambda$ can be complex
if $\mid E-V_{n} \mid <$ 2: this
corresponds to extended states. If $\mid E-V_{n} \mid >$ 2, then $\lambda$
is real, corresponding to localized states.
There is a metal-insulator transition at the so--called mobility edge
$\mid E-V_{n} \mid=$2.

In order to determine the actual nature of the spectrum for
irrational $\omega$, a barrage of mathematical techniques have been
applied to the study of the Harper equation \cite{Harper,kspla,sk,prss,nr,ks}. 
One of the main results has been the discovery of {\it duality}
by Andr\'e and Aubry \cite{aubry}.  Since the Harper potential contains 
no harmonics, there can be a solution to Eq.~(\ref{harpq}) such that 
\beq
\psi_{n}= \exp\{ikn\} \sum_{m=-\infty}^\infty f_{m} \exp\{im(2\pi\omega 
n+\phi_0)\}  
\eqn
where the Fourier coefficients $f_m$ are given by
\beq
f_{m}= \exp\{-ikm\} \sum_{m=-\infty}^\infty \psi_{n} \exp\{-in(2\pi\omega m
+\phi_0)\}.
\eqn
Substituting the value of $\psi_{n}$ in the above equation yields
\beq
f_{m+1} + f_{m-1} + \frac{2}{\epsilon} \cos (2\pi\omega m+k) f_{m} = \frac{1}{\epsilon} E f_{m}
\label{rpq},
\eqn
which, written as
\beq
f_{m+1} + f_{m-1} + 2\tilde{\epsilon} \cos (2\pi\omega m+k) f_{m} = \tilde{E} f_{m}
\eqn 
is similar in form to Eq.~(\ref{harpq}), with $\tilde{\epsilon} 
\equiv 1/\epsilon$ and $\tilde{E} \equiv E/\epsilon$. The equations 
are {\it identical} when $$ \epsilon = 1, \phi_0  = k$$
with $ \psi \equiv  f.$
Now if $f_{m}$ is localized then the sum $\sum_{m=-\infty}^\infty 
\mid f_{m} \mid^2$ is finite, and therefore the function
\beq
f(x)= \exp\{ikx\} \sum_{m=-\infty}^\infty f_{m} \exp\{im2\pi\omega x\}
\eqn
converges. Thus $f(x)$ behaves like a Bloch function and represents
an extended state.  For $ x = n + \frac{\phi_0}{2\pi\omega}$, 
$f(x)= \exp\{ik \frac{\phi_{0}}{2 \pi\omega}\} \psi_{n}$. 
Hence if $\psi_{n}$ is extended then $\sum_{n}\mid \psi_{n}\mid^2$ 
diverges. Therefore, if $\psi_{n}$ is extended then $f_{n}$ is
localized. 

The duality of the model is this correspondence which maps the region
$\epsilon >$ 1 to the region $\epsilon <$ 1, with the point $\epsilon = 1$
being ``self dual''. Thus states for this value of $\epsilon$ are neither
extended nor localized: they are {\it critical}.

This criticality has two manifestations. The spectrum of the Harper
equation at $\epsilon =$ 1 is singular continuous: the
eigenvalues form a Cantor set. The wave--functions for these critical
states are power--law localized. 
As a function of $\omega$, the spectrum of the Harper equation, which
was first studied in some detail by Hofstadter \cite{hofstadter} has
a remarkable shape. For every rational value, it consists of a finite
number of bands, while for irrational values of $\omega$ it consists
of an apparently fractal set of points, giving what is termed the
Hofstadter butterfly.

A number of other quasiperiodic potentials
\cite{kohmoto,pandit,kohmoto2,luck} have since been shown to support 
critical states. One widely studied system is the Kohmoto model, 
defined at lattice site $n$ through the rule \cite{kohmoto,pandit}
\beqr
V_n &=& \alpha~~~~~~~~~~~~~ 0 \le \{n\omega\} \le \omega \nonumber\\ &=&
-\alpha~~~~~~~~~~~\omega < \{n\omega\} \le 1,
\label{fibo}
\end{eqnarray}
(the notation is $\{y\} \equiv y$ mod 1,) which is quasiperiodic
if $\omega$ is an irrational number. For the case of $\omega =
\gamma = (\sqrt 5 -1 )/2$, the golden mean ratio, one obtains the
so--called Fibonacci chain.

This latter potential is the subject of the present paper.
Our approach is based on the equivalence between the
discrete Schr\"odinger equation and a derived iterative mapping for
the amplitude ratio of the wave function at neighbouring sites. This
equivalence was first noted by Bondeson \etl \cite{bondeson} who showed
that the Schr\"odinger equation with a quasiperiodic potential could
be transformed into a dynamical system with quasiperiodic forcing.
Ketoja and Satija \cite{sk} extended this analysis to the Harper
equation, Eq.~(\ref{harpq}), obtaining the entirely equivalent 
Harper map,
\begin{eqnarray}
\label{ks}x_{k+1} &=& {-1 \over{x_k - E + 2 \epsilon \cos 2 \pi
\phi_k}}\\ 
\phi_{k+1} &=& \{\phi_k + \omega\}, \label{ks2} 
\end{eqnarray}
using the transformation $\psi_{i-1}/\psi_i \to x_i$. This is a
skew--product driven iterative mapping of the infinite strip
$(\-\infty,\infty) \otimes [0,1]$ to itself.  If the frequency
$\omega$ is an irrational number, the driving is quasiperiodic in
time. For the Kohmoto model, the analogous corresponding map is
\beq
\label{ksf}x_{k+1} = {-1 \over{x_k - E + V_k}}
\eqn
with $V_k$ given by Eq.~(\ref{fibo}).
In either case, 
the quantum problem is meaningful only when $E$ is an eigenvalue, but
in the above map, $E$ appears only as a parameter. Such driven maps,
which were first introduced by Grebogi \etl \cite{gopy}, have been
studied extensively in the context of strange nonchaotic dynamics
\cite{rmp}. In quasiperiodically driven dissipative systems, the
dynamics can (for appropriate parameter values) be on fractal sets
which concurrently have non-positive Lyapunov exponents: such
motion is on strange nonchaotic attractors (SNAs).

As has now been demonstrated in several studies, localized states
correspond to SNAs \cite{sk,prss,nr}. The transition from extended to
localized states as a function of potential strength can be viewed as
a transition to SNAs. Critical states have a special
significance: these are SNAs on which {\em all} Lyapunov exponents
are zero.  In this paper we use this fact to study the states of the
Fibonacci chain for which it is known explicitly that all states are
critical, by determining the eigenvalue spectrum of the system
through the condition that the Lyapunov exponent for the specified
value of $E$ be zero. This provides a simpler alternative than the
transfer matrix techniques, and in addition, gives a characterization
of the states of the system which is not accessible to purely quantum
mechanical methods \cite{nr}.

In the following section we briefly review the relevant features of
SNAs and the methods used to study them. Section III contains the
main results of this work for the Fibonacci chain. 
Section IV contains a discussion of related potentials
deriving from abstract aperiodic sequences, followed by a brief summary.

\section {Strange Nonchaotic attractors}

Strange nonchaotic attractors are frequently found in nonlinear
systems where the forcing is quasiperiodic \cite{rmp}. There are
similarities between SNAs and both periodic as well as chaotic
attractors.  Like the former, they are characterized by zero or
nonpositive Lyapunov exponents, and like the latter, they have a
fractal structure \cite{gopy}.  Owing to fractal structure of SNAs,
the dynamics is strictly aperiodic.  Numerous examples are known now
of systems with SNAs. These include, in addition to discrete
quasiperiodically forced maps, a number of continuous dynamical
systems such as driven pendulums and oscillators \cite{rmp,venkat}.

In the mapping first discussed by Grebogi, Ott, Pelikan and Yorke
\cite{gopy} 
\begin{eqnarray}
\label{tanh}
x_{i+1} &=& 2 \alpha \cos 2\pi \theta_i \tanh x_i\\
\theta_{i+1} &=& \{\omega+ \theta_i\}.
\label{irra}
\end{eqnarray}
the reasoning that establishes the existence of strange nonchaotic
motion is as follows. For $\omega$ an
irrational number, there are no periodic orbits in this system.  The
mapping $x\to 2\alpha \tanh x$ is 1--1 and contracting, taking the
real line into the interval $[-2\alpha ,2\alpha ]$. The dynamics in
$\theta $ is ergodic in the unit interval and therefore the attractor 
of the dynamical system must be contained in the strip $[-2\alpha,
2\alpha ]\otimes [0,1]$. This region contains an invariant subspace,
namely the line $x = 0$.  A point $x_n$ with the corresponding 
$\phi _n=$ 1/4 will map to ($x_{n+1}=0,\theta_{n+1}=\omega +1/4$), and 
hence the subsequent iterates will all remain in this invariant 
subspace.  At the same time, the line $x=0$ can be made unstable by 
increasing $\alpha$: the transverse Lyapunov exponent is $\lambda =\ln 
\vert \alpha\vert$. Therefore, for $\vert \alpha\vert > 1$ the 
transverse Lyapunov exponent is positive and the line $x=0$ is no
longer attracting. The total Lyapunov exponent can, however, become negative
for sufficiently large $\alpha$ so there is an attractor which lies in the
region $x=\pm 2\alpha$, with a dense set of points on the line $x=0$,
as well as points off this line. The above arguments can be put on
firm mathematical footing to show that the attractor is both strange
and nonchaotic \cite{keller} for $\alpha > 1$.
For the Harper map, similar arguments \cite{ks,plethora} can be
advanced to suggest that for large coupling the dynamics is on SNAs.

In systems such as the quasiperiodically driven circle map
\cite{sosno} or the logistic map \cite{pmr}, 
\beqr
\label{logistic}
x_{n+1} &=& \alpha [1 + \epsilon \cos 2\pi \theta_n] x_n (1 - x_n) \\
\theta_{n+1}  &=&  \{\theta_n + \omega \},
\eqnr
SNAs are frequently observed in the neighborhood of the transition to
chaos \cite{sosno,pmr}, so it is important to determine the
scenarios through which they can be formed, as well as the different
methods used for their characterization.  To date there are several
routes known for the creation of SNAs; these have been reviewed
recently \cite{rmp}.  SNAs can be characterized by calculating the
Lyapunov exponents, fractal dimensions, as well as correlation
functions and related quantities \cite{pf}.

For map of Harper type, namely of the form Eq.~(\ref{ks}-\ref{ks2}) or 
Eq.~(\ref{ksf}), the nontrivial Lyapunov exponent can be easily  computed
as
\beq
\label{lyap}
\lambda = \lim_{N \to \infty}\sum_{j=0}^{N-1} \ln x_{i+1}^2
\eqn
while the other Lyapunov exponent is trivially zero. Note that since
the dynamics is invertible, there can be no chaos in this system.

\section{The Fibonacci Chain}
\label{secfibo}

The Fibonacci chain is the simplest example of a quasicrystal in 
one-dimension, and can be thought of as a two-component lattice. 
Using the two symbols $a$ and $b$, with initial sequences 
$S_{0}=b, S_{1}=a$, the recursive substitution $a \rightarrow a b, 
b \rightarrow a$ gives 
\beq
S_{l+1} = \{ S_{l} \cdot S_{l-1} \}
\eqn
with $\{\cdot\}$ representing concatenation.
It is clear that the sequence $S_{k}$  has $F_{k}$ elements, 
where $F_{k}$ is a Fibonacci number specified by 
$F_{k+1}=F_{k}+ F_{k-1}$, $F_{0}=1, F_{1}=1$. Labeling sites
of a lattice of length $F_{k}$ by the element of a sequence $S_{k}$,
and specifying the site potential $V_{l}= \alpha$ if the symbol is $a$,
$V_{l}= - \alpha$ if the symbol is $b$ gives the Fibonacci 
chain of length $F_{k}$. We are interested in the eigenvalue spectrum
of this chain in the limit $k \to \infty$.

The Fibonacci chain is the particular case of $\omega = \gamma \equiv
(\sqrt 5 -1 )/2$, in the more general Kohmoto model, Eq.~(\ref{fibo}).
For the Fibonacci chain, and more generally for any 
irrational $\omega$ in the Kohmoto model, it is known that 
all states are critical for any $\alpha$ \cite{kohmoto,pandit}. 
We can therefore use the alternate `quantization condition', 
$\{ E: \lambda(E,\omega,\alpha) = 0\}$ to obtain the eigenvalue
spectrum of the Kohmoto model, as shown in Fig.~1 for the value $\alpha = 1$.
The Cantor set structure which is also evident in the spectrum
for all irrational $\omega$ is more clearly seen in the dependence 
of $\lambda$ on $E$ at fixed $ \omega, \alpha,$ as in Fig.~2.
The curve $\lambda(E)$ meets the line $\lambda=0$ on a Cantor set of points,
namely the spectrum, and in the spectral gaps, the curve is 
parabolic in shape. 

The gaps of the spectrum can be studied for any irrational $\omega$;
we consider here the case of the golden mean since this
case has been considered in detail earlier. The behavior of 
gap widths as a function of $\alpha$ has been of interest, not just
in the Kohmoto model, but in the Harper equation \cite{nr3} and other
related models potentials derived from aperiodic sequence as 
well \cite{luck,jagannathan}.

Note that since the Lyapunov exponent is strictly nonpositive in this system,
the dynamics is {\it entirely} on attractors (strange as well as non--strange).
Following Johnson and Moser \cite{john} it is possible to define a winding
number as a function of $E$ as \cite{brazil}
\beq
\Omega (E) =\lim_{N \to \infty} {1 \over {2 N}} W(x_1,\ldots,x_N)
\eqn
where $W(x_1,\ldots,x_N)$ is an indicator function that counts the
number of changes of sign in the orbit $x_1,\ldots,x_N$. Since the 
variable $x_k$ is the ratio of the wavefunction at neighbouring sites,
$\Omega(E)$ essentially counts the number of nodes in the wavefunction
per unit length, and
is therefore the normalized integrated density of states.
The winding number is shown as a function of $E$ for $\alpha =1$ in the
Fibonacci chain in Fig.~3; within a spectral gap, the winding number 
which remains constant, can be expressed as $p+q\gamma$, with $p,q$ integers
\cite{numbther}. This provides a labeling of the gaps which
are indicated for a few of the larger gaps in Fig.~3.
  
For fixed $\omega$, the ordering of the individual eigenstates do not 
change with the potential strength $\alpha$. However gap widths $w$ vary 
with $\alpha$ in a complicated manner. The width of the largest gap, 
namely the one marked $A$ in Fig.~2, scales linearly: $w_A \sim \alpha$.
Those marked $B$ and $C$ asymptote to constant width, $w_B, w_C \sim \alpha^0$,
while all other gaps decrease in width, but as power, $w \sim \alpha^{-\mu}$,
the exponent $\mu$ depending on the particular gap. The dependence
of width on $\alpha$ is shown for gaps $A-D$ in Fig.~4.
In the case of Harper equation it has been shown \cite{nr} that each gap can be 
indexed by distinct topological integer index, which determines the 
scaling of the widths. Scaling results are also available for the 
Thue--Morse and period--doubling potentials \cite{luck}. 
In the present case it has not been possible to deduce any simple
number--theoretic dependence of the gap exponents as in the Harper
case, though clearly, here too the gaps never close since
the spectrum is always singular continuous. 

The depths characterising the gaps, namely the minimum value that the 
LE takes inside a gap also vary with $\alpha$ in a similar fashion, 
increasing for $A$, asymptoting to constant for $B$ and $C$, and 
decreasing as a power for all other gaps.

\section{Summary and Discussion}
\label{conclu}

The equivalence between critical SNAs and critically localized states
provides a new means of determining the spectrum of 
discrete quasiperiodic systems. We have, in previous studies \cite{nr,nr3},
extensively applied this technique to the study of the Harper
equation, and here we make application to the case of the Kohmoto model,
specifically the case of the Fibonacci chain,
where the potential takes values $\pm \alpha$ in a
quasiperiodic manner as described in Section \ref{secfibo}. 

Similar dichotomous
potentials which have been studied in the context of critical
localization are the Thue-Morse, period-doubling and Rudin-Shapiro 
sequences \cite{luck}. The methods employed here, namely the derivation
of an iterative mapping equivalent to the discrete Schr\"odinger equation
and the identification of critically localized states with critical
SNAs can be applied to this entire class of problems. 

The present method makes it possible to empirically study the eigenvalue
spectrum of such systems in considerable detail. As in related
quasiperiodic problems \cite{luck,nr}, the scaling of the gap widths with 
potential strength $\alpha$ depends on the particular gap being
considered. For the Fibonacci chain the scaling is $w \sim \alpha^{\mu}$,
with the exponent $\mu = 1$ for the largest gap, $\mu = 0$ for the next
two, and $\mu \le -1$ for the remaining gaps.  The organization of the 
gaps, which can be uniquely labeled by two integers, remains 
incompletely understood, but it is likely that, as in the Harper 
problem \cite{nr}, the exponent governing the power--law decay of 
gap widths will be related to the gap labels. These aspects, and the 
application of the present technique to more general aperiodic
sequences is the subject of ongoing work \cite{negithesis}.

\section*{Acknowledgment} This work is supported by a grant from the
Department of Science and Technology.

\newpage
\centerline{Figure Captions}

\begin{figure}[!htb]
\centerline{\def\epsfsize#1#2{0.8#1}\epsffile{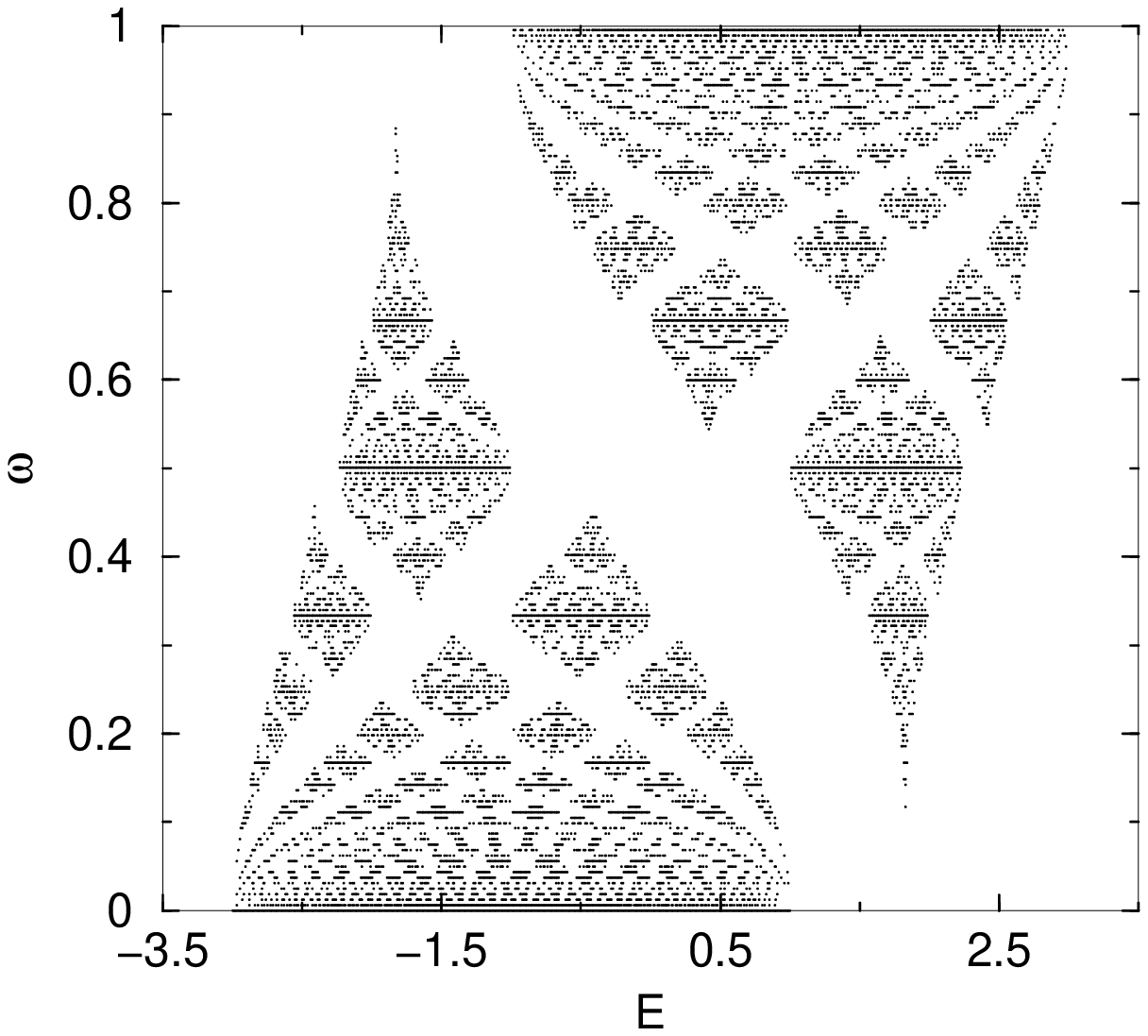}}
\caption{\label{fig1.eps}
Phase diagram for the Kohmoto model for $\alpha = 1$.}
\end{figure}

\begin{figure}[!htb]
\centerline{\def\epsfsize#1#2{0.8#1}\epsffile{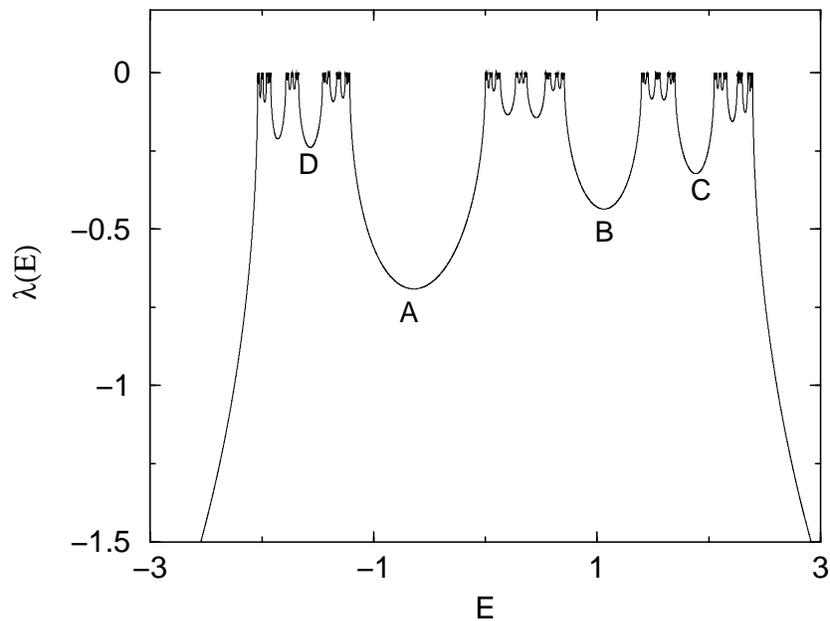}}
\caption{\label{fig2.eps}
Lyapunov exponent versus energy at $\alpha=1$ for the Fibonacci
chain, namely $\omega = (\sqrt 5 -1)/2.$ The largest visible gaps 
are labelled $A, B, C$ and $D$ respectively.
The dynamics for $\lambda = 0$ corresponds to SNAs.}
\end{figure}

\begin{figure}[!htb]
\centerline{\def\epsfsize#1#2{0.8#1}\epsffile{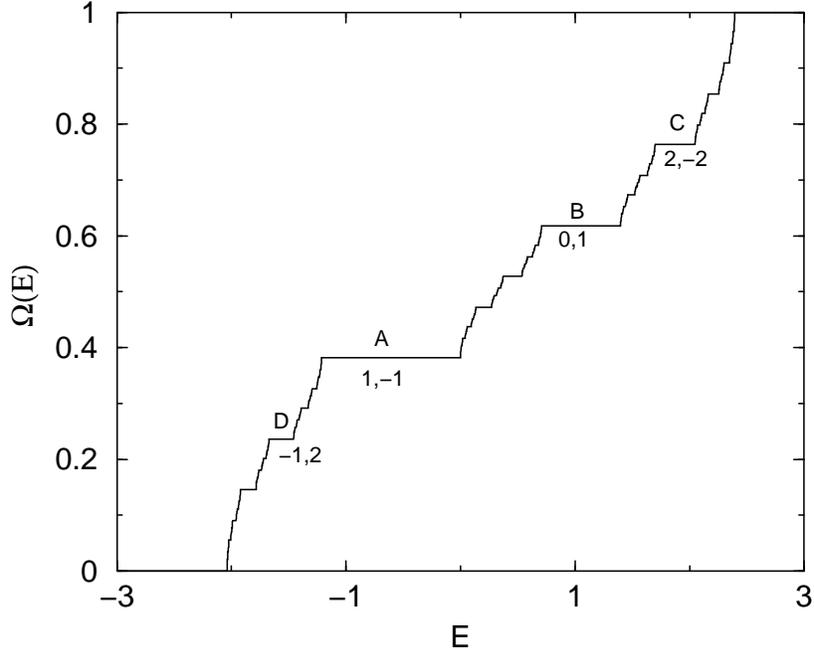}}
\caption{\label{fig3.eps}
Integrated density of states as a function of energy $E$ for
$\omega = \gamma = (\sqrt 5 -1)/2$ at $\alpha=1.$
On the plateaux corresponding to the gaps $A, B, C$ and $D$ the
rotation number can be expressed as $p+q\gamma$, with $p,q$ 
integers. These are indicated for the largest gaps.}
\end{figure}

\begin{figure}[!htb]
\centerline{\def\epsfsize#1#2{0.8#1}\epsffile{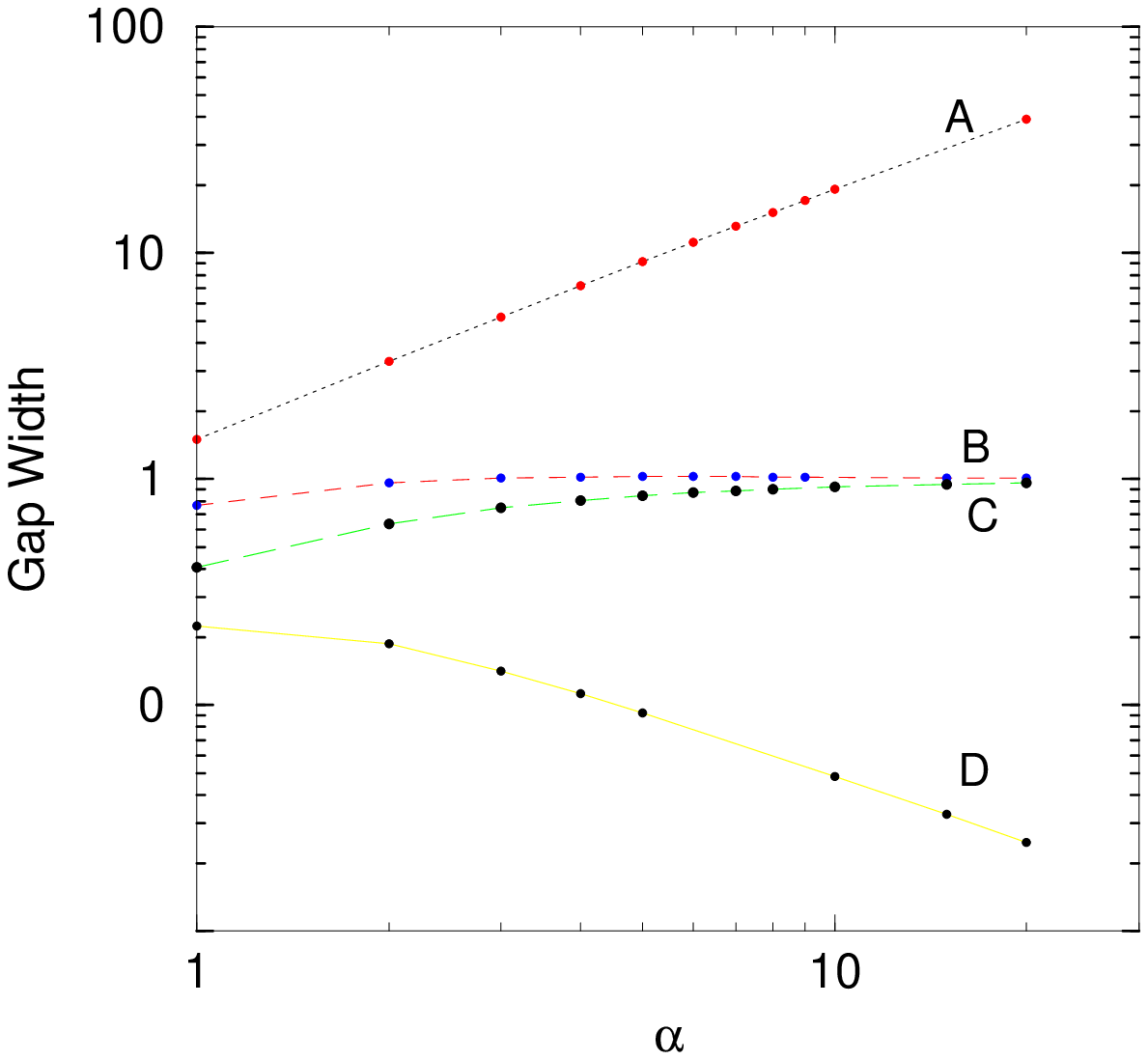}}
\caption{\label{fig4.eps}
Scaling of the gap widths, $w$ in the Fibonacci chain for the 
largest gaps as a function of $\alpha$.}
\end{figure}


\begin{thebibliography}{99}

\bibitem{anderson} P.W. Anderson, \pr{109}{1492}{1958}; \rmp{50}{191}{1978}.
\bibitem{bds} B. L. Altshuler and B. D. Simon, {\it Universalities: From Anderson
localization to quantum chaos}, Elsevier Science Publisher, B. V. {1994}.
\bibitem{hike} M. Hike, \jpha{30}{L367}{1997}.
\bibitem{mott-twose} N. F. Mott,  {\it Metal-Insulator Transitions}, (Taylor
and Francis, London 1990)
\bibitem{Azbel} M. Ya. Azbel, Sov. Phys. JETP, {\bf 19}, 634 (1964);
\prl{43}{1954}{1979}.
\bibitem{soko} J. B. Sokoloff, \prp{126}{189}{1985}.
\bibitem{thpb} D. J. Thouless, \prb{28}{4272}{1983}.
\bibitem{aubry} G. Andr\'e and S. Aubry, Ann. Isr. Phys. Soc., {\bf 3},
133 (1980).
\bibitem{fib11}M. Kohmoto and J. Banavar, \prb{34}{563}{1986}.
\bibitem{fib12}Y. Liu, W. Sritrakool, \prb{43}{1110}{1991}.
\bibitem{fib13}G. Gumbs and M. K. Ali, \prl{60}{1081}{1998}.
\bibitem{fib14}M. Kohmoto, B. Sutherland and C. Tang, \prb{35}{1020}{1987}.
\bibitem{Harper} P. G. Harper, \prlon{68}{874}{1955}.
\bibitem{thouless} D. J. Thouless, \jpcs {5}{77}{1972}.
\bibitem{barelli} A. Barelli, J. Bellissard, P. Jacquod and D. L. Shepelyansky \prl{23}{4752}{1996}.
\bibitem{kohmoto} M. Kohmoto, \prl{51}{1198}{1983}.
\bibitem{pandit} S. Ostlund, R. Pandit, D. Rand, H. J. Schellnhuber,
and E. D. Siggia, \prl{50}{1873}{1983}.
\bibitem{kohmoto2} M. Kohmoto, L. P. Kadanoff, and C. Tang, \prl{50}{1870}{1983}.
\bibitem{luck} J. M. Luck, \prb{39}{5834}{1989}.
\bibitem{ag} A. G. Abanov, J. C. Talstra, P. B. Wiegmann, \prl{81}{2112}{1998}.
\bibitem{gisel} T. Geisel, R. Ketzmerick, G. Petschel, \prl{66}{1651}{1991}.
\bibitem{sinai} Ya. G. Sinai, \jsph{46}{861}{1987}.
\bibitem{simon} R. del Rio, S. Jitomirskaya, Y. Last, and B. Simon,
\prl{75}{117}{1995}.
\bibitem{bell} J. Bellisard, R. Lima, and D. Testard,
\cmp{88}{107}{1983}.
\bibitem{john} R. Johnson and J. Moser, \cmp{84}{403}{1982}.
\bibitem{numbther} M. Waldschmidt, P. Moussa, J. -M. Luck and C. Itzykson {\it 
From Number Theory to Physics,} Springer-Verlag, 1989.
\bibitem{stockmann} H.J. Stockmann, {\it Quantum Chaos : An Introduction}, (Cambridge University
Press, Cambridge, 1999).
\bibitem{kspla}  J. Ketoja and I. Satija, \pla{194}{64}{1994}.
\bibitem{sk} J. Ketoja and I. Satija, \prl{75}{2762}{1995}.
\bibitem{prss} A. Prasad, R. Ramaswamy, I. Satija, and N. Shah, \prl{83}{4530}{1999}.
\bibitem{nr} S. S. Negi and R. Ramaswamy, Phys. Rev. E, submitted.
\bibitem{ks}  J. Ketoja and I. Satija, \physd{109}{70}{1997}.
\prl{83}{4530}{1999}.
\bibitem{hofstadter} D. R. Hofstadter, \prb{14}{2239}{1976}.
\bibitem{bondeson} A. Bondeson, E. Ott, and T. M. Antonsen, \prl{55}{2103}{1985}.
\bibitem{gopy} C. Grebogi, E. Ott, S. Pelikan, and J. Yorke, \physd{13}{261}{1984}.
\bibitem{rmp} A. Prasad, S. S. Negi, and R.  Ramaswamy, Int. J. Bifurcation and Chaos, to be published, 2001.
\bibitem{venkat}A Venkatesan, M Lakshmanan,A Prasad and R Ramaswamy, \pre{61}{3641}{2000}.
\bibitem{keller} G. Keller, \fund {151}{139}{1996}.
\bibitem{plethora} S. S. Negi and R. Ramaswamy, \pram{56}{47}{2001}.
\bibitem{sosno} O. Sosnovtseva, U. Feudel, J. Kurths, and A. Pikovsky, \pla{218}{255}{1996}.
\bibitem{pmr} A. Prasad, V. Mehra and R. Ramaswamy, \pre{57}{1576}{1998}.
\bibitem{pf} A. Pikovsky and U. Feudel, Chaos, {\bf 5}, 253 (1995).
\bibitem{nr3} S. S. Negi and R. Ramaswamy, Phys. Rev. E, submitted.
\bibitem{jagannathan} F. Piechon, M. Benakli and A. Jagannathan, 1995 
\lanl, cond--mat/9502068.

\bibitem{brazil} P. C. Ferreira, F. P. Mancini and M. H. R. Tragtenberg, 2000
\lanl, cond--mat/0002329.

\bibitem{negithesis} S S Negi, Ph D thesis, Jawaharlal Nehru University, 
New Delhi, 2001.

\end{thebibliography}
\end{document}